%% file: ms_final.tex
\documentclass{emulateapj}
\usepackage{graphics}
\usepackage{color}
\usepackage{amsmath}
\usepackage{empheq} 
\newcommand{\ltsimeq}{\la}
\newcommand{\gtsimeq}{\ga}
\newcommand{\lsun}{L$_{\odot}$}
\newcommand{\msun}{M$_{\odot}$}

\newcommand{\HI}{H{\sc i}}

\accepted{September 2016}
\shortauthors{McQuinn et al.}
\shorttitle{The Distance to M104}

\begin{document}
\title{The Distance to M104}\thanks{Based on observations made with the NASA/ESA Hubble Space Telescope, obtained from the Data Archive at the Space Telescope Science Institute, which is operated by the Association of Universities for Research in Astronomy, Inc., under NASA contract NAS 5-26555.}
\author{Kristen.~B.~W. McQuinn\altaffilmark{1,2}, 
Evan D. Skillman\altaffilmark{2},
Andrew E.~Dolphin\altaffilmark{3},
Danielle Berg\altaffilmark{4}, 
Robert Kennicutt\altaffilmark{5}
}

\altaffiltext{1}{University of Texas at Austin, McDonald Observatory, 2515 Speedway, Stop C1400 Austin, Texas 78712, USA \ {\it kmcquinn@astro.as.utexas.edu}}
\altaffiltext{2}{Minnesota Institute for Astrophysics, School of Physics and Astronomy, 116 Church Street, S.E., University of Minnesota, Minneapolis, MN 55455, USA} 
\altaffiltext{3}{Raytheon Company, 1151 E. Hermans Road, Tucson, AZ 85756, USA}
\altaffiltext{4}{Center for Gravitation, Cosmology and Astrophysics, Department of Physics, University of Wisconsin Milwaukee, 1900 East Kenwood Boulevard, Milwaukee, WI 53211, USA}
\altaffiltext{5}{Institute for Astronomy, University of Cambridge, Madingley Road, Cambridge CB3 0HA, England }

\begin{abstract}
M104 (NGC~4594; the Sombrero galaxy) is a nearby, well-studied elliptical galaxy included in scores of surveys focused on understanding the details of galaxy evolution. Despite the importance of observations of M104, a consensus distance has not yet been established. Here, we use newly obtained Hubble Space Telescope optical imaging to measure the distance to M104 based on the tip of the red giant branch method. Our measurement yields the distance to M104 to be $9.55\pm0.13 \pm0.31$ Mpc  equivalent to a distance modulus of $29.90 \pm 0.03 \pm 0.07$ mag. Our distance is an improvement over previous results as we use a well-calibrated, stable distance indicator, precision photometry in a optimally selected field of view, and a Bayesian Maximum Likelihood technique that reduces measurement uncertainties. The most discrepant previous results are due to Tully-Fisher method distances, which are likely inappropriate for M104 given its peculiar morphology and structure. Our results are part of a larger program to measure accurate distances to a sample of well-known spiral galaxies (including M51, M74, and M63) using the tip of the red giant branch method. 
\end{abstract} 

\keywords{galaxies:\ spiral -- galaxies:\ distances and redshifts -- stars:\ Hertzsprung-Russell diagram}

\section{Introduction}\label{sec:intro}
\subsection{M104; The Sombrero Galaxy}

\begin{figure*}
\includegraphics[width=\textwidth]{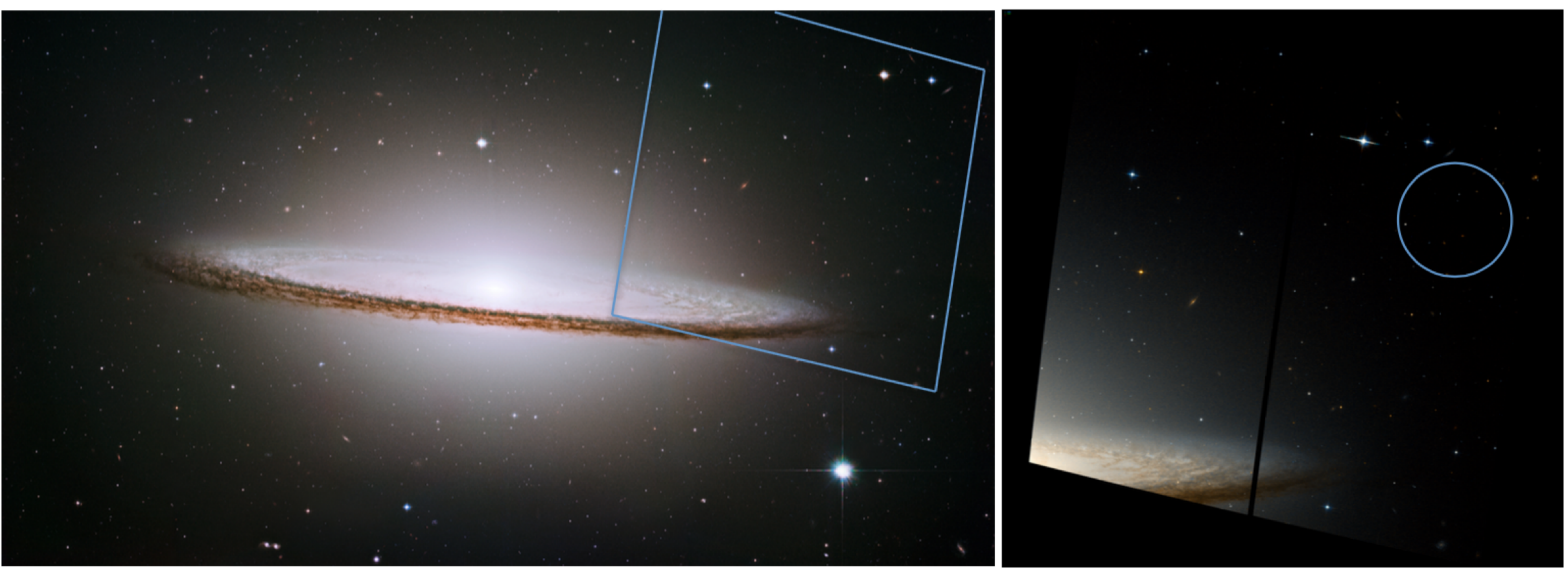}
\caption{{\textit Left:} HST Heritage composite image of M104 from NASA/ESA and The Hubble Heritage Team STScI/AURA created using $F435W$, $F555W$, and $F625W$ observations with the footprint of our new HST observations overlaid. {\textit Right:} HST ACS imaging of our new field created by combining F606W (blue), F814W (red), and an average of the two filters (green). The field selected for our TRGB measurements is overlaid in blue at an average galactocentric radius of $4.\arcmin6$. The images are oriented with North up and East left.}
\label{fig:image}
\end{figure*}

M104 (NGC~4594; the Sombrero galaxy) is a peculiar elliptical galaxy that is widely recognized from its striking sombrero hat-like appearance (see Figure~\ref{fig:image}). The system was originally classified as a large-bulge spiral galaxy based on the prominent disk of gas, dust, and stars seen in optical images. However, when viewed in {\it Spitzer Space Telescope} infrared imaging, the disk is seen to be enveloped in a significantly larger, approximately spherical distribution of old stars. Based on its infrared morphology, M104 was re-classified as an elliptical galaxy \citep{Gadotti2012}. Although still debated, these authors propose that the disk of material in the inner region of the galaxy has been accreted. M104 hosts a central, supermassive black hole ($M \sim10^9$ \msun), based on spectroscopically determined stellar velocities \citep{Kormendy1988, Kormendy1996}, and is classified as a low ionization nuclear emission region galaxy \citep[LINER;][]{Ho1997}. 

In addition to individual studies, M104 has also been extensively studied across the electromagnetic spectrum and included in many detailed surveys of nearby galaxies including the Spitzer Infrared Nearby Galaxies Survey \citep[SINGS;][]{Kennicutt2003}, the GALEX Space Telescope Nearby Galaxy Survey \citep[NGS;][]{GildePaz2007}, and the Key Insights on Nearby Galaxies: a Far-Infrared Survey with Herschel program \citep[KINGFISH;][]{Kennicutt2011}. 

Surprisingly, M104 lacks a secure distance measurement.  Measured distances vary widely, ranging from 12.6$-$21.7 Mpc from the Tully-Fisher (TF) relation (although see \S\ref{sec:comparison} for a discussion of applicability of the TF method on such a peculiar galaxy), 7.8$-$9.08 Mpc from the planetary nebula luminosity function (PNLF) method, 8.55$-$10.0 Mpc from surface brightness fluctuations (SBF), 6.22$-$15.80 Mpc from the globular cluster luminosity function (GCLF) method, and one preliminary tip of the red giant branch (TRGB) distance measurement of 10.7 Mpc that was not adopted in subsequent analysis (discussion and references in Section~\ref{sec:comparison} and see, also, the Appendix). As a result, the derived properties of M104 that depend fundamentally on distance are based on approximate distances spanning $6.2-21.7$ Mpc. Many of the physical and derived quantities (stellar mass, gas mass, and luminosity based star-formation rates, to name a few) depend on the square of the distance. Thus, results across studies built on this wide range of distances are difficult to compare and may be subject to large systematic offsets.

\subsection{Precise TRGB Distances}
We have undertaken a survey to measure precise distances using the TRGB method to well-studied nearby galaxies that lack secure distances. The focus of this paper is M104. In \citet[][hereafter Paper I]{McQuinn2016a} we reported the distance to M51 (the Whirlpool galaxy) to be $8.58\pm0.10$ (statistical) Mpc, and we described the observation strategy, data analysis, and methodology for the full program. Our sample also includes the Sunflower (M63; NGC~5055), M74 (NGC~628; the archetype grand-design spiral), and 4 additional spiral galaxies from the SINGS program (NGC~5398, NGC~1291, NGC~4559, NGC~4625). 

The TRGB luminosity in the $I$-band is an accurate indicator of the distance to a galaxy and has been applied to numerous samples of galaxies in the nearby universe. Originally compared to the accuracy of Cepheid distances, the TRGB method is now preferred over Cepheid-based distances as the Cepheid period-luminosity relation may differ from galaxy to galaxy \citep[e.g.,][]{Tammann2008, Mould2008, Ngeow2012}. The theoretically well-understood TRGB distance methodology is based on the predictable luminosity of low-mass stars immediately prior to the helium flash \citep{Mould1986, Freedman1988, DaCosta1990, Lee1993}. Low-mass red giant branch (RGB) stars increase in luminosity while burning hydrogen in a shell of material with an outer convective envelope. Independent of stellar mass, the helium in the electron degenerate core ignites, resulting in a He-flash. This core helium burning phase begins at a predictable luminosity, with only a small dependency on stellar metallicity in the $I$-band \citep[e.g.,][]{Lee1993, Salaris1997}. In addition to being well-understood and predictable, an advantage of the TRGB method is the TRGB luminosity is well-calibrated to the Hubble Space Telescope (HST) filters, with empirically determined corrections for the modest metallicity dependence \citep{Rizzi2007a}. 

The TRGB distance method requires imaging of resolved stellar populations in the $I$-band reaching $\sim1$ mag below the TRGB (for identifying the discontinuity of the TRGB) and the V band (for selecting RGB stars from composite stellar populations). This can be achieved via single orbit per filter HST observations for galaxies within the Local Volume, making the TRGB method not only precise but also efficient for nearby galaxies. 

Here for M104, we measure the TRGB distance employing the same approach from Paper~I and report our results similarly. In Sections~\ref{sec:data} and \ref{sec:distances}, we briefly summarize the observations, data analysis, and present the distance measurement. For a more complete description, we refer the interested reader to Paper~I. In Section~\ref{sec:comparison}, we compare our distance value with other measurements from the literature. Our conclusions are summarized in Section~\ref{sec:conclusions}. In the Appendix of our study of M51, we provided descriptions of various previous 
methods used to measure the distance to the galaxy. Many of those same methods have been used to measure the distance to M104. For brevity, we do not repeat the descriptions here, but add an additional method (the GCLF) used for M104 in our Appendix.

\input{tab1}

\begin{figure*}
\includegraphics[width=\linewidth]{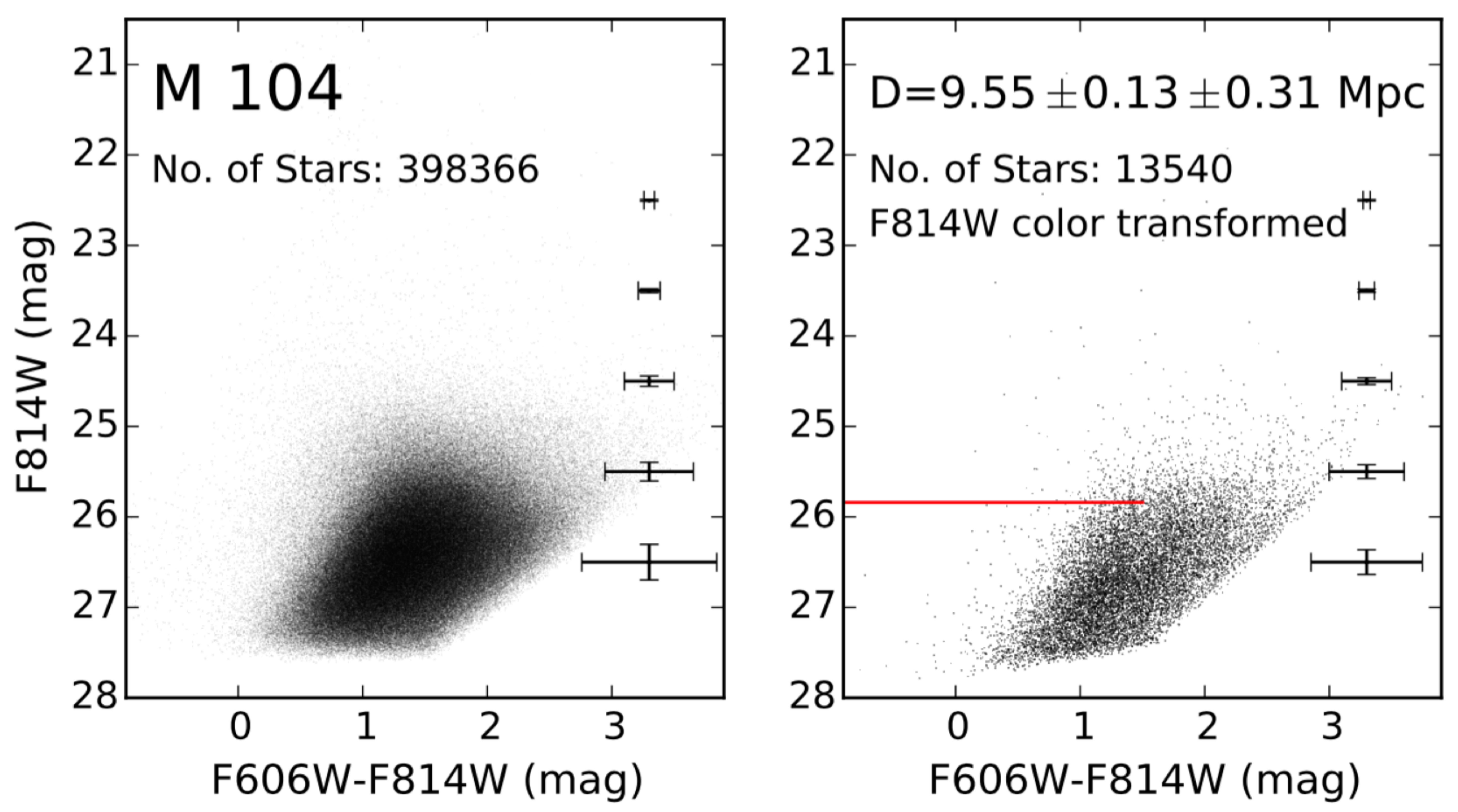}
\caption{{\em Left:} Color-magnitude diagram (CMD) of the full HST field of view. {\em Right:} CMD of the field of view selected for TRGB analysis. Both CMDs have been corrected for foreground extinction. The uncertainties include uncertainties from the photometry and artificial star tests. The uncertainties in color are larger due to the inclusion of F606W photometry with low SNR. The measured TRGB is marked with a horizontal red line. The photometry in the right panel was transformed using the color-based calibration correction for metallicity. By applying this correction before fitting for the TRGB (instead of applying the correction in the final calibration), the curvature in the RGB is reduced allowing for the TRGB to be measured with a higher degree of certainty.}
\label{fig:cmd}
\end{figure*}

\section{Observations and Photometry}\label{sec:data}
Table~\ref{tab:properties} lists the coordinates, foreground extinction, and observation details for M104. The observations were obtained as part of the HST-GO-13804 program (PI: McQuinn) with the HST using the Advanced Camera for Surveys (ACS) Wide Field Channel (WFC) \citep{Ford1998}. In an identical manner to the M51 observations, the images include 1 orbit observations in each of the F606W and F814W filters using a 2-point hot pixel dither pattern with integration times of $\sim2500$ s. 

The angular extent of M104 is significantly larger than the ACS field of view. Thus, we carefully selected a field of view for our observations that avoided the crowded center of M104 (making high fidelity photometry difficult), while still located at a small enough radius that a sufficient number of RGB stars are within the ACS field of view. Selecting a field in the outer parts of the galaxy has the additional benefits of (i) reducing the contribution of younger asymptotic giant branch (AGB) stars that can complicate the identification of the TRGB and (ii) lowering the average metallicity of the RGB stars which ensures a bluer color of the TRGB and proper photometric depth in the F606W filter is achieved. More details on determining the field selection for observational planning purposes can be found in Paper~I, Section~2.

The left panel of Figure~\ref{fig:image} presents an HST Heritage\footnote{M104 was observed for 18 orbits in a 6 field mosaic with the ACS in 2003 in order to make the Hubble Heritage image, but observations were obtained in F435W, F555W, and F625W only.  Observations in F814W were required for a TRGB distance measurement.} image of M104 overlaid with the ACS field of view of our new observations. The right panel of Figure~\ref{fig:image} presents the new HST imaging obtained for the TRGB measurement created by combining the F606W (blue), the average of the F606W and F814W images (green), and F814W (red) observations. These images were made using the CTE corrected images (flc.fits files) for each filter and combined with \textsc{Astrodrizzle} from DrizzlePac 2.0. The image is dominated by the higher surface brightness region of the disk, but the lower surface brightness outer region is also well-populated with stars. 

The data reduction and processing were done in an identical manner to the processing of the M51 data. We repeat a summary of the process here from Paper~I. Photometry was performed on the images after processing by the standard ACS pipeline and included charge transfer efficiency (CTE) corrections to reduce the impact of non-linearities caused by space radiation damage on the ACS instrument \citep[e.g.,][]{Anderson2010, Massey2010}. We used the photometric software DOLPHOT\footnote{URL: http://americano.dolphinsim.com/dolphot/}, a modified version of HSTphot optimized for the ACS instrument \citep[][]{Dolphin2000}. The photometry was filtered to include well-recovered, high-fidelity sources based on a number of measured parameters for each point source, including signal-to-noise ratios (SNRs), and sharpness and crowding parameters. As the F814W magnitudes are used in the TRGB measurement, we applied a minimum SNR of 5$\sigma$, ensuring higher significance photometric measurements in the distance determination. As the F606W magnitudes are only used to provide color constraints for selecting RGBs stars, we applied a less stringent SNR minimum of 2$\sigma$. Sources were culled based on the sharpness and crowding parameters in order to avoid sources whose PSF are sharply peaked or broad, or whose photometric uncertainty is higher due to the proximity of other sources. Completeness limits of the images were measured by performing artificial star tests with DOLPHOT and filtering the output on the same parameters as the photometry. The photometry was corrected for Galactic extinction based on the dust maps of \citet{Schlegel1998} with updated calibration from \citet{Schlafly2011}; these values are provided in Table~\ref{tab:properties}.

The left panel in Figure~\ref{fig:cmd} shows the extinction corrected color-magnitude diagram (CMD) from the full ACS field of view. Photometric depth in the CMD corresponds to the 50\% completeness level determined from the artificial star tests. Representative uncertainties per magnitude from the PSF fitting photometry and artificial star tests are also shown. 
The CMD is primarily populated by a clearly defined RGB sequence with photometry reaching $\sim2$ mag below the approximate TRGB identifiable by eye. 

As reported in Figure~\ref{fig:cmd}, the photometry from the full ACS field of view recovers nearly 400k stars. Given the range in structure and crowding, and the large number of stars within the field of view, we chose to apply spatial cuts to the photometry. The judicious application of spatial cuts minimizes photometric uncertainties by focusing on measurements from less crowded regions in the outer parts of the galaxy with a potentially smaller relative number of AGB stars. Thus, we chose to avoid the disk region of M104, selecting instead a region $\sim1.\arcmin7$ above the plane for our TRGB measurement with an average galactocentric radius of $4.\arcmin6$; the selected field is highlighted in the right panel of Figure~\ref{fig:image}. The photometric completeness in the magnitude range of the TRGB increased from an average of $\sim$70\% in the full field of view to $\gtsimeq$85\% in our selected region, as measured by artificial star tests. This final region used for the TRGB measurement includes $\sim15$k stars. 

The TRGB magnitude has a modest dependency on metallicity which can be taken into account when calibrating the measured luminosity of the TRGB to a distance modulus. The metallicity correction is color-based and typically uses the average TRGB $F606W - F814W$ color to account for the difference between the target galaxy and the average $V-I$ TRGB color of 1.6 used in the calibration. For convenience, we reproduce the relation from \citet{Rizzi2007a}:

\begin{equation}
M_{F814W}^{ACS} = -4.06+ 0.20 \cdot [(F606W-F814W) - 1.23] \label{eq:trgb}
\end{equation}

\noindent Instead of applying the metallicity correction after measuring the TRGB luminosity, we apply this color-based correction to the photometry {\em prior} to fitting for the TRGB, thereby reducing the curvature and width of the RGB and increasing the sharpness in the break of the luminosity function (LF). In the right panel of Figure~\ref{fig:cmd}, we present the extinction corrected CMD from the region selected for the TRGB analysis after applying the color-based correction for metallicity. We use these data to fit for the break in the LF corresponding to the TRGB. The final zeropoint for the TRGB based distance is applied to the final measurement. 

\begin{figure}
\includegraphics[width=1.08\linewidth]{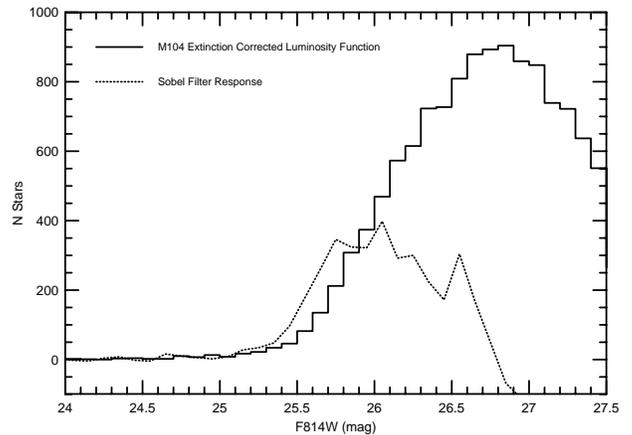}
\caption{F814W luminosity function of the stars with F606W-F814W colors between 0.5$-$2.3 mag in the selected region of M104 used in the distance measurement. The photometry has been transformed with the color-based metallicity and corrected for Galactic extinction. The Sobel filter response is overlaid on the LF.}
\label{fig:lf_sobel}
\end{figure}

\section{Distance Determination from the TRGB}\label{sec:distances}
We used a Sobel filter edge detection technique with a Sobel kernel in the form [$-2$, 0, $+2$] \citep{Lee1993, Sakai1996, Sakai1997} and a Bayesian Maximum Likelihood (ML) technique \citep[see ][for a full discussion]{Makarov2006} to identify and measure the discontinuity in the F814W LF corresponding to the TRGB. The ML technique does not rely on binning the LF and takes into account photometric error distribution and completeness from artificial star tests. As a result, the ML technique typically has lower measured statistical uncertainties and is therefore an improvement over the Sobel filter approach. However, the Sobel filter approach is useful to provide a consistency check on the best-fitting ML result. Both methods are discussed in more detail in Paper~I, including the theoretical LF used in the ML technique (see Equations~1a,b). To minimize the number of non-RGB stars in the data that can add Poisson noise to the TRGB fit, we selected sources in the $F606W - F814W$ color range of $0.5-2.3$, consistent with colors of RGB stars. 

Figure~\ref{fig:lf_sobel} shows the F814W LF for the sources from the CMD in right panel of Figure~\ref{fig:cmd}, with the Sobel filter response overlaid. The first Sobel peak is at 25.75$^{-0.1}_{+0.2}$ mag; we assign a larger uncertainty on the fainter end given that the peak response extends into the adjacent fainter magnitude bin. As seen in the CMD, the shape of the Sobel peak reflects the more slowly changing density of sources at the bright end of the RGB population. The best-fitting value for the extinction-corrected TRGB luminosity from the ML technique is F814W $= 25.84\pm0.02$, in agreement with the discontinuity identified by the Sobel filter. We adopt the ML TRGB measurement in our final distance calculations. The identified TRGB is noted in Figure~\ref{fig:cmd} in the CMD from the outer disk with the metallicity correction applied. 

Using the measured TRGB magnitude, we apply only the zero-point calibration from Equation~\ref{eq:trgb} \citep{Rizzi2007a} to calculate a distance modulus of 29.90 $\pm$ 0.03 mag corresponding to a distance to M104 of 9.55 $\pm$ 0.13 Mpc (statistical). Uncertainties are based on adding in quadrature the statistical uncertainties from the TRGB zero-point calibration ($\sigma=$ 0.02), the color-dependent metallicity correction ($\sigma=$ 0.01), and the statistical ML uncertainties calculated from the probability distribution function, which include uncertainties from the photometry and artificial star tests. The final numbers are also noted in Table~\ref{tab:distances}.

As discussed in Paper~I, the systematic uncertainties for the TRGB calibration are not well-quantified. The TRGB calibration we use from \citet{Rizzi2007a} is anchored by the horizontal branch calibration of \citet{Carretta2000} who use an averaged distance to the Large Magellanic Cloud from various methods as it distance base. \citet{Carretta2000} report $1\sigma$ combined statistical and systematic uncertainties on their zeropoint calibration of 0.07 and on their metallicity correction of 0.025 at the metallicity of the sample examined in \citet{Rizzi2007a}. Combined, these systematic uncertainties are equivalent to 0.074 mag. This estimate of the systematic uncertainty is slightly lower than the 0.12 reported in \citet{Bellazzini2001} based on a TRGB calibration using the globular cluster $\omega$Cen and direct distance estimates from a detached eclipsing binary. Here, we choose to adopt the systematic uncertainties from \citet{Carretta2000} which is the basis for our adopted calibration in \citet{Rizzi2007a}, but note the systematic uncertainties on the distance modulus may be slightly higher and of order 0.12 mag. Including systematic uncertainties, our final distance to M104 is 9.55 $\pm 0.13 \pm 0.31$ Mpc.

\input{tab2}

\section{Comparison with Previous Distances}\label{sec:comparison}
\begin{figure}
\includegraphics[width=\linewidth]{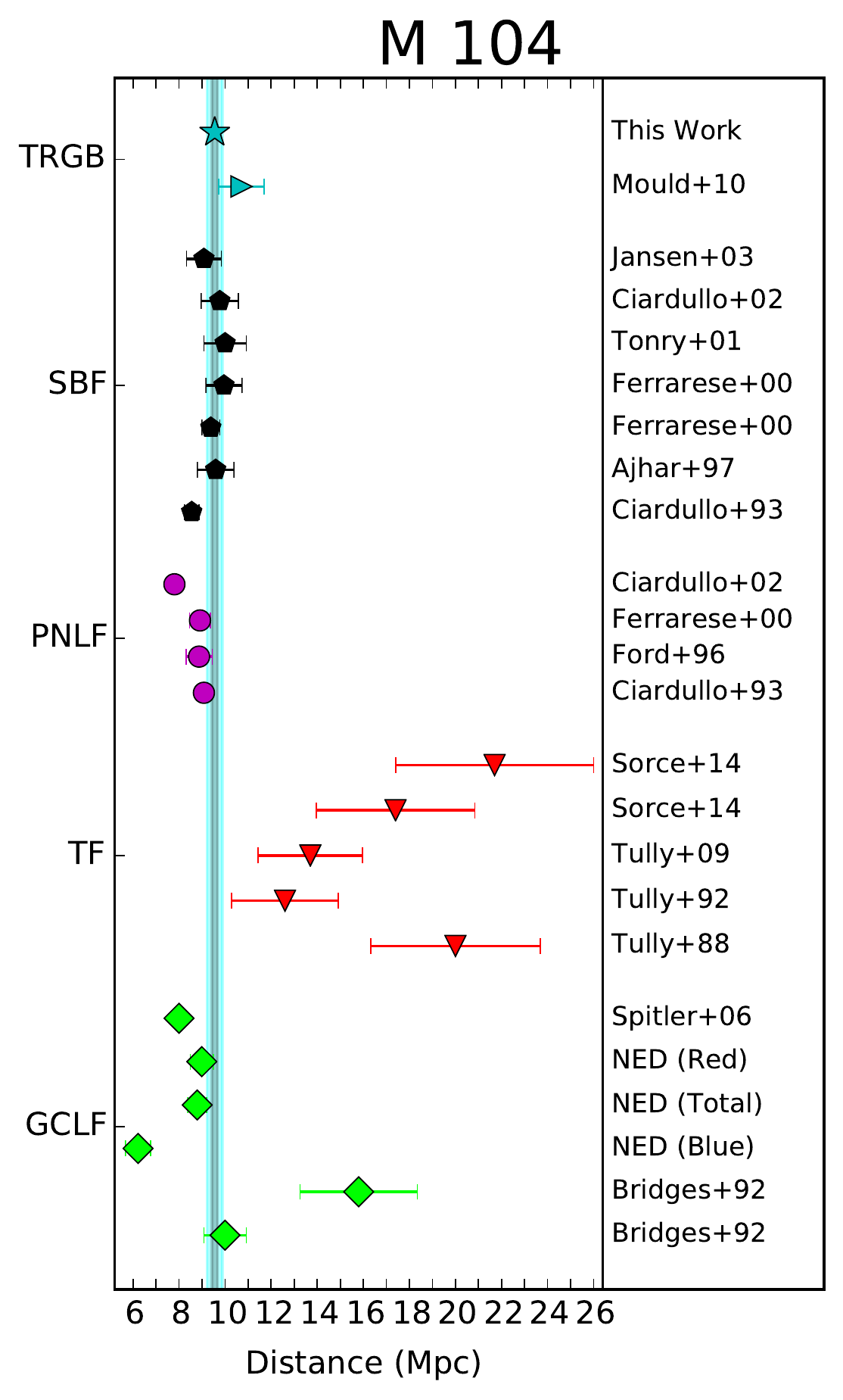}
\caption{Comparison of distance measurements to M104 from the literature. To aid the comparison between our value and the previous values, we have added a shaded vertical line in grey (cyan) centered on our TRGB measurements whose width encompasses the 1~$\sigma$ statistical (systematic) uncertainty on our measurement. See Table~\ref{tab:distances} for individual distance values, references, and sources for the data.}
\label{fig:comparison}
\end{figure}

Figure~\ref{fig:comparison} compares our TRGB distance measurement to M104 with 23 other reported distance measurements using various techniques from 15 sources from the literature. The distances were compiled from individual measurements listed in the NASA/IPAC Extragalactic Database (NED). To aid in the comparison with previous results, we overlay two vertical shaded lines centered on our distance measurement with widths encompassing the 1$\sigma$ statistical and systematic uncertainties in distance. The individual values, methods, and references for the different distances in Figure~\ref{fig:comparison} are listed in Table~\ref{tab:distances}; a comparative discussion of our result with the different methods is presented below. The SBF, PNLF, and TF methods were also used to measure the distance to M51 in Paper~I, where brief descriptions of these techniques are provided in the Appendix. An additional technique, the GCLF method, has been used to measure the distance to M104. For completeness, we provide a summary of the GCLF approach in the Appendix of this work. 

In Figure~\ref{fig:comparison}, the various distance measurements are grouped by technique and, within each technique, are listed from the most recent to the oldest. Some of the studies measured multiple distances for M104 \citep[i.e.,][]{Bridges1992, Ciardullo1993, Ferrarese2000, Sorce2014}, also detailed in Table~\ref{tab:distances}. We include the separate, individual measurements to show the range in distances derived by the different studies. These multiple measurements were used to calibrated distance indicators across methods \citep[e.g.,][]{Ciardullo1993, Ferrarese2000} or to compare/recalibrate a single method \citep[e.g.,][]{Bridges1992, Sorce2014}. Note that some studies used archival data from previous publications and report revised distances based on recalibration or different data reduction techniques. Thus, in Table~\ref{tab:distances} we list not only the reference for the distance, but also the source of the data. Interestingly, NED independently calculates three distances based on the GCLF method using the data listed in \citet{Larsen2001} which had adopted the PNLF distance from \citet{Ford1996} for their analysis. In this case, we cite NED as the source of the distance with observations from \citet{Larsen2001}.

The previous distance measurements to M104 cover a wide range of values, from $6.22 - 21.70$ Mpc. Our distance measurement lies towards the lower end of the range, where the majority of measurements from the SBF, PNLF, and GCLF are clustered. The TF results return a higher range in distances, with some overlap from the GCLF method. 

\citet{Mould2010} report a TRGB distance modulus of 30.15$\pm0.2$ (corresponding to a distance of 10.7$\pm0.1$ Mpc) from archival HST WFPC2 imaging located at a galactocentric radius of $\sim10\arcmin$. The CMD reaches $\ltsimeq1$ mag below the TRGB. Few details are provided on the measurement of the TRGB, however the authors reference the study of \citet{Mould2008} as the basis for their calculation. Subsequent analysis in the study does not use this TRGB distance, but instead adopts the SBF distance from \citet{Tonry2001}. 

The SBF distances span a $\sim1.5$ Mpc range from seven separate measurements, six of which overlap with our TRGB distance measurement at $1\sigma$. The SBF approach uses degradations in the resolution of galaxy as a function of distance as its fundamental basis. Smoothly varying, older stellar populations provide the most stability for the SBF method. Because the SBF method uses all RGB stars observed in a CMD, the calibration requires a steep dependency on metallicity. The SBF method has typical reported uncertainties of $\pm1$ Mpc; systematic and calibration uncertainties have been difficult to quantify \citep{Ferrarese2000}.

The PNLF distance uncertainties overlap with our reported distance for three of the four measurements. All four measurements use the same data set from \citet{Ford1996}. In Figure~\ref{fig:comparison}, the reported uncertainties are smaller than the plot symbol for three of the four measurements. However, similar to the SBF distances, there are a number of unquantified uncertainties in the PNLF distances including a possible dependence of the calibration on the age of the central star in the PN and extinction corrections of the inherently dusty region around AGB stars in a metal-rich galaxy. The PNLF method also shows a possible weak dependence on the galaxy metallicity that is not included in the calibration, but which likely has a smaller impact for higher metallicity systems such as M104 \citep{Ciardullo2002}. 

The TF distances include the original TF distance to M104 and four additional measurements using a combination of archival data for the \HI\ line widths as well as photometric data in the optical and infrared. All of the TF values are offset to larger distances; none overlap with our TRGB distance measurement. We examined whether there were other peaks in the Sobel filter responses out to the limit of our data that might correspond to these farther distances. We identified additional peaks at $\sim26.05$ and $\sim26.55$ mag, corresponding to $\sim10.5$ and $\sim13.2$ Mpc respectively, which are clearly below the TRGB identifiable in Figure~\ref{fig:cmd}. The larger TF distances in Figure~\ref{fig:comparison} are unphysical given the constraints from the CMD. 

Although, historically, M104 has been classified as an Sa galaxy \citep{deVaucouleurs1991},
the more recent study by \citet{Gadotti2012} has proposed that M104 has more of the 
characteristics of an elliptical galaxy.  These authors suggest that M104 formed as an 
elliptical galaxy but accreted a massive disc, with a contemporary result as a ``peculiar 
system''.  

Regardless of whether or not M104 is an elliptical galaxy with an accreted disc 
or a peculiar large-bulge spiral, the properties of M104 are highly unusual. 
Looking at 21 cm neutral hydrogen observations in the literature, the \HI\ radial distribution extends
only as far as the dust ring seen in Figure~\ref{fig:image}, with a rotation velocity of 
350 km s$^{-1}$ in the flattened part of the rotation curve \citep{Faber1977}. 
The measured 21 cm neutral hydrogen line width at 20\% has an exceptionally high 
maximum width of $\sim$760 km s$^{-1}$.
This line width is larger than any used in the calibration of the TF relation.
Based on the $M_{HI}$ measurement of $2.8\times10^8$ \msun\ and a 
$L_B$ of $2.4\times10^{10}$ \lsun\ from \citet{Bajaja1984}, adjusted using our distance,
we calculate an $M_{HI}/L_B$ ratio of 0.01 \msun/\lsun, 
a value significantly lower than in normal spiral galaxies. 

Given these properties, it is not surprising that the TF method of measuring distances
produces a discrepant result.  The TF method is dependent on a relationship between the
luminosity (or mass) and maximum rotation speed for spiral galaxies, and thus should not
be expected to produce a meaningful distance estimate for a peculiar galaxy (elliptical or spiral)
with extreme properties. Similarly, there is a known offset in the TF relation 
for S0 galaxies \citep{Williams2010} in the sense that S0 galaxies are less luminous 
at a given rotational velocity, which is in the same sense as observed for M104.

The GCLF method yields a range of distances from $6.2 - 15.8$ Mpc; two of which overlap with our TRGB distance. M104 has one of the largest numbers of globular clusters cataloged in a nearby galaxy, estimated to be $1900\pm200$ \citep{Hargis2014}. The color distribution of the globular clusters shows bi-modality, suggesting metal-poor and metal-rich subpopulations. \citet{Larsen2001} measure the turn-over luminosity of the blue, the red, and the combined populations; NED uses these measurements with a zeropoint also reported in \citet{Larsen2001} to calculate and report three distances. Because these distances are not reported in \citet{Larsen2001}, we cite NED as the reference. Given the uncertain history of the disk in M104, it is possible that a non-negligible fraction of the globular cluster population may have been accreted to M104. In this case, the globular cluster population may be a larger mix or systems formed around M104 and systems accreted, which adds uncertainty to this distance method.

\section{Conclusions}\label{sec:conclusions}
Using the stable and well-calibrated TRGB method and $HST$ optical imaging of resolved stellar populations, we measure the distance modulus and distance to M104, the Sombrero galaxy, to be $29.90\pm0.03\pm0.07$ and $9.55\pm0.13\pm0.31$ Mpc respectively. We adopt the systematic uncertainty on the TRGB calibration from \citet{Carretta2000}, but note that it may be slightly higher and of order 0.12 mag \citep{Bellazzini2001}. The TRGB in the CMD was identified using a ML technique which takes into account photometric uncertainties and incompleteness in the data, and converted to a distance using the calibrations of \citet{Rizzi2007a} specific to the $HST$ filters with a metallicity correction. 

Previously reported distances for M104 range from $6.22 - 21.70$ Mpc; the majority of measurements cluster between $8 - 11$ Mpc with outliers to greater distances from the PNLF, TF and GCLF methods. The TF method is likely not applicable for M104 given its peculiar gas distribution and high velocity, and uncertain morphological classification. Our measurement is an improvement over previous distance measurements as we use (i) the well-understood TRGB distance indicator that is a stable and predictable standard candle, (ii) precision calibration of the method that includes second order corrections, (iii) precise photometry in an uncrowded field with careful application of both spatial and color cuts to the data, and (iv) a robust Bayesian ML technique to measure the TRGB feature in the data, which includes measurements of photometric incompleteness and does not rely on binning the LF. 

\vspace{18pt}
Support for this work was provided by NASA through grant GO-13804 from the Space Telescope Institute, which is operated by Aura, Inc., under NASA contract NAS5-26555. This research made use of NASA's Astrophysical Data System and the NASA/IPAC Extragalactic Database (NED) which is operated by the Jet Propulsion Laboratory, California Institute of Technology, under contract with the National Aeronautics and Space Administration. Finally, the authors thank the anonymous referee for constructive comments which helped improve this work. 

\appendix
In Paper~I \citep{McQuinn2016a}, we gave brief descriptions of the different methods used to measure the distance to M51. These included three of the methods found in the literature used to measure the distance to M104 including Surface Brightness Fluctuations (SBF), Planetary Nebulae Luminosity Functions (PNLF), and the Tully-Fisher relation (TF). We refer the interested reader to the Appendix of Paper~I for details on these methods. An additional metric was also used to measure the distance to M104, namely the Globular Cluster Luminosity Function (GCLF). Here, we briefly describe this additional method.

\section{Globular Cluster Luminosity Function (GCLF) Distances}\label{sec:appendix}
The luminosity function of the composite globular cluster system of individual galaxies has been noted empirically to turn-over at approximately the same absolute magnitude (calibrated using other methods such as Cepheids, PNLF, SBF, or TRGB based distances). Thus, by observing a large number of globular clusters down to a completeness level well below this peak magnitude, the GCLF of a galaxy can be used as a standard candle distance approach. The GCLF approach implies there is a characteristic mass function for globular clusters, which has yet to be understood. While many GCs are thought to form at early times coincident with the formation of the host galaxies, GCs are also thought to be accreted to the host galaxy via hierarchical merging of lower-mass galaxies with their own GCs. This stochastic process could add significant uncertainties to parameterizing a `universal' GCLF. Statistical uncertainties in the method are introduced by background galaxies mis-identified as globular clusters, difficult to quantify incompleteness as a function of galactic radius, and uncertain reddening corrections. Using complete samples of globular clusters around a galaxy can help reduce uncertainties due to small number statistics, but can also introduce uncertainties as the spread in age and metallicity of the clusters can impact the turn-over magnitude. Using samples of metal-poor globular clusters can help mitigate degeneracies introduced by the age and metallicity changes, but can also increase uncertainties due to inherently smaller sample sizes. 

\renewcommand\bibname{{References}}
\bibliography{../../bibliography.bib}

\end{document}

%% file: tab1.tex
\begin{table}
\begin{center}
\caption{M104 Properties and Observations}
\label{tab:properties}
\end{center}
\begin{center}
\vspace{-15pt}
\begin{tabular}{ll}
\hline 
\hline 
Parameter				& Value	\\
\hline
RA (J2000)			& $12:39:59.4$ \\
Dec (J2000)			& $-11:37:23$	\\
$A_{F606W}$			& 0.126 mag \\
$A_{F814W}$			& 0.078 mag \\
$F606W$ exp. time		& 2549 s \\
$F814W$ exp. time		& 2549 s \\

\hline
\\
\end{tabular}
\end{center}
\vspace{-10pt}
\tablecomments{Observation times are from program GO$-$13804 (PI McQuinn). Galactic extinction estimates are from \citet{Schlafly2011}.}

\end{table}

%% file: tab2.tex
\begin{table*}
\begin{center}
\caption{Distance Measurements to M104}
\label{tab:distances}
\end{center}
\begin{center}
\vspace{-15pt}
\begin{tabular}{llll}
\hline 
\\
dm (mag)				& D (Mpc)	& Reference			& Data	\\
\hline 
\hline
\\
\multicolumn{4}{c}{\textbf{Tip of the Red Giant Branch (TRGB)}} \\
29.90$\pm0.03\pm0.07$		& $9.55\pm0.13\pm0.31$ & \textbf{This work} & new observations \\
\\
30.15	$\pm0.2$		& 10.71	& \citet{Mould2010}		& archival \\
\\
\multicolumn{4}{c}{\textbf{Surface Brightness Fluctuations (SBF)}} \\
29.79	$\pm$0.18	& 9.08	& \citet{Jensen2003} 	& new observations \\ 
29.91	$\pm$0.18	& 9.59 	& \citet{Ciardullo2002} 	& \citet{Tonry2001} \\ 
29.95 	$\pm$0.18	& 9.77	& \citet{Tonry2001} 		& new observations \\  
30.01	$\pm$0.20	& 10.00	& \citet{Ferrarese2000}	& \citet{Tonry2001, Ajhar2001} \\  
29.99	$\pm$0.17	& 9.95	& \citet{Ferrarese2000} 	& \citet{Tonry2001, Ajhar2001} \\
29.86	$\pm$0.09	& 9.38	& \citet{Ajhar1997} 		&  \citet{Lauer1997} \\ 
29.66	$\pm$0.08	& 8.55	&  \citet{Ciardullo1993}	&  \citet{Ford1996} \\  
\\
\multicolumn{4}{c}{\textbf{Planetary Nebulae Luminosity Function (PNLF)}} \\ 
29.46	$\pm$0.08	 & 7.80	& \citet{Ciardullo2002} 	& \citet{Ford1996} \\ 
29.75	$\pm$0.11		 & 8.91	& \citet{Ferrarese2000} 	& \citet{Ford1996} \\ 
29.74	$\pm$0.14	 & 8.87	& \citet{Ford1996} 		& new observations \\ 
29.79	$\pm$0.07	 & 9.08	& \citet{Ciardullo1993} 	& \citet{Ford1996} \\ 
\\
\multicolumn{4}{c}{\textbf{Tully-Fisher Relation (TF)}}\\
31.68	$\pm$0.43	& 21.70	&\citet{Sorce2014} 		& archival \\ 
31.20	$\pm$0.43	& 17.40	& \citet{Sorce2014} 		& archival \\ 
30.69	$\pm$0.36	& 13.70	& \citet{Tully2009} 		& archival \\ 
30.50	$\pm$0.40	& 12.60	& \citet{Tully1992} 		& archival \\ 
31.50	$\pm$0.40	& 20.00	& \citet{Tully1988} 		& archival \\ 
\\
\multicolumn{4}{c}{\textbf{Globular Cluster Luminosity Function (GCLF)}}\\ 
29.52	$\pm$0.06	& 8.00	& \citet{Spitler2006}		& archival \\
29.77	$\pm$0.12	& 8.99	& NASA/IPAC ED (NED)	& \citet{Larsen2001}  \\
29.72	$\pm$0.10	& 8.79	& NASA/IPAC ED (NED)	& \citet{Larsen2001} \\
28.97	$\pm$0.19	& 6.22	& NASA/IPAC ED (NED)	& \citet{Larsen2001} \\
31.00	$\pm$0.35	& 15.80	& \citet{Bridges1992}		& new observations \\
30.00	$\pm$0.20	& 10.00	& \citet{Bridges1992}		& new observations \\
\\
\hline\hline
\end{tabular}
\end{center}
\tablecomments{Distance measurements from the literature from various techniques. The Reference column lists the source of the reported measurement. The Data column lists whether the observations were original to the study, from data archives, or a re-calibration of existing work in the literature. Figure~\ref{fig:comparison} shows the distribution of the measurements.  Details on the SBF, PNLF, and TF methods can be found in the Appendix of Paper~I; details on the GCLF method can be found in the Appendix of this work.}

\end{table*}